\documentclass[oldversion]{aa}

\usepackage{txfonts}
\usepackage{natbib}
\bibpunct{(}{)}{;}{a}{}{,}
\usepackage{graphicx}
\usepackage{rotating}

\begin{document}

\defcitealias{P2005002}{Paper~I}
\defcitealias{A2005015}{T05}

\title{Eclipsing binaries suitable for distance determination in the \\
Andromeda galaxy\thanks{Based on observations made with the Isaac Newton
Telescope operated on the island of La Palma by the Isaac Newton Group in the
Spanish Observatorio del Roque de los Muchachos of the Instituto de
Astrof\'{\i}sica de Canarias.}\fnmsep\thanks{Tables 3--5 are available in
electronic form at the CDS via anonymous ftp to \texttt{cdsarc.u-strasbg.fr
(130.79.128.5)} or via
\texttt{http://cdsweb.u-strasbg.fr/cgi-bin/qcat?J/A+A/XXX/XXX}}}

\author{F. Vilardell\inst{1} 
        \and I. Ribas\inst{2,3} 
        \and C. Jordi\inst{1,3}}

\offprints{F. Vilardell, \email{fvilarde@am.ub.es}}

\institute{Departament d'Astronomia i Meteorologia, Universitat de 
Barcelona, c/ Mart\'{\i} i Franqu\`es, 1--11, 08028 Barcelona, Spain\\
 \email{[fvilarde;carme.jordi]@am.ub.es}
 \and
  Institut de Ci\`encies de l'Espai--CSIC, Campus UAB, Facultat de 
Ci\`encies, Torre C5-parell-2a, 08193 Bellaterra, Spain\\
 \email{iribas@ieec.uab.es}
 \and 
  Institut d'Estudis Espacials de Catalunya (IEEC), Edif. Nexus, c/ Gran
Capit\`a, 2-4, 08034 Barcelona, Spain 
}

\date{Received \today{} / Accepted \today{}} 

\abstract
{The Local Group galaxies constitute a fundamental step in the definition of
cosmic distance scale. Therefore, obtaining accurate distance determinations to
the galaxies in the Local Group, and notably to the Andromeda Galaxy (M31), is
essential to determining the age and evolution of the Universe. With this
ultimate goal in mind, we started a project to use eclipsing binaries as
distance indicators to M31. Eclipsing binaries have been proved to yield direct
and precise distances that are essentially assumption free. To do so,
high-quality photometric and spectroscopic data are needed. As a first step in
the project, broad band photometry (in Johnson $B$ and $V$) has been obtained
in a region ($34^\prime\times 34^\prime$) at the North--Eastern quadrant of the
galaxy over 5 years. The data, containing more than 250 observations per
filter, have been reduced by means of the so-called difference image analysis
technique and the DAOPHOT program. A catalog with 236\,238 objects with
photometry in both $B$ and $V$ passbands has been obtained. The catalog is the
deepest ($V<25.5$ mag) obtained so far in the studied region and contains
3\,964 identified variable stars, with 437 eclipsing binaries and 416 Cepheids.
The most suitable eclipsing binary candidates for distance determination have
been selected according to their brightness and from the modelling of the
obtained light curves. The resulting sample includes 24 targets with
photometric errors around 0.01 mag. Detailed analysis (including spectroscopy)
of some 5--10 of these eclipsing systems should result in a distance
determination to M31 with a relative uncertainty of 2--3\% and essentially free
from systematic errors, thus representing the most accurate and reliable
determination to date.}

 \keywords{Stars: variables: general -- Stars: binaries: eclipsing -- Stars:
fundamental parameters -- Stars: distances -- Techniques: photometric --
Catalogs} 

\maketitle

\section{Introduction}

Extragalactic distance determinations depend largely on the calibration of
several distance indicators (e.g., Cepheids, supernovae, star clusters, etc) on
galaxies of the Local Group with known distances. The Large Magellanic Cloud
(LMC) has traditionally been used as a f\mbox{}irst rung for extragalactic distance
determinations. However, even taking into account that some recent results seem
to converge to an LMC distance modulus of $(m-M)_0=18.50\pm0.02$
\citep{A2005024}, its low metallicity and a possible line-of-sight extension
have posed serious difficulties to the accurate calibration of several distance
indicators. The Andromeda galaxy (M31), on the contrary, has a more simple
geometry and a metallicity more similar to the Milky Way and other galaxies
used for distance estimation \citep[see, e.g.,][]{A2006005}. Therefore,
although stars in M31 are about six magnitudes fainter than those in LMC, the
particular characteristics of this galaxy make it a promising f\mbox{}irst step of the
cosmic distance scale \citep{A2006008}.

Given the importance of M31 as an anchor for the extragalactic distance scale,
many studies have provided distance determinations to M31 using a wide range of
methods. A comprehensive list of distance determinations to M31, with explicit
errors, are shown in Table \ref{distances}. As can be seen, the values listed
are in the range $(m-M)_0=24.0-24.6$ mag. Although the weighted standard
deviation is of $\sim$4\%, most of the distance determinations in Table
\ref{distances} rely on previous calibrations using stars in the Milky Way or
the Magellanic Clouds. As a consequence of this, a large number of subsequent
distance determinations, based only on new recalibrations, can be found in the
literature. These are not included in Table \ref{distances}.  Therefore, a
direct and precise distance determination to M31 is of central importance since
this would permit the use of all the stellar populations in the galaxy as
standard candles.

\begin{table}[!ht]
 \caption{Distance determinations to M31 as presented in the references. Values
resulting of posterior calibrations and distance moduli without extinction
corrections are not included.}
 \label{distances}
 \centering
 \begin{tabular}{l r@{$\pm$}l r@{$\pm$}l c}
  \hline
  \hline
  Method & \multicolumn{2}{ c }{$(m-M)_0$} & \multicolumn{2}{ c }{Distance} & Reference \\
  & \multicolumn{2}{ c }{[mag]} & \multicolumn{2}{ c }{[kpc]} & \\
  \hline
  Cepheids & 24.20 & 0.14 & 690 & 40 & [1] \\
  Tip of the RGB & 24.40 & 0.25 & 760 & 90 & [2] \\
  Cepheids & 24.26 & 0.08 & 710 & 30 & [3] \\
  RR Lyrae & 24.34 & 0.15 & 740 & 50 & [4] \\
  Novae & 24.27 & 0.20 & 710 & 70 & [5] \\
  Cepheids & 24.33 & 0.12 & 730 & 40 & [6] \\
  Cepheids & 24.41 & 0.09 & 760 & 30 & [6] \\
  Cepheids & 24.58 & 0.12 & 820 & 50 & [6] \\
  Carbon--rich stars & 24.45 & 0.15 & 780 & 50 & [7] \\
  Cepheids & 24.38 & 0.05 & 752 & 17 & [8] \\
  Carbon--rich stars & 24.36 & 0.03 & 745 & 10 & [8] \\
  Glob. Clus. Lum. Func. & 24.03 & 0.23 & 640 & 70 & [9] \\
  Red Giant Branch & 24.47 & 0.07 & 780 & 30 & [10] \\
  Red Clump & 24.47 & 0.06 & 780 & 20 & [11] \\
  Red Giant Branch & 24.47 & 0.12 & 780 & 40 & [12] \\
  Cepheids & 24.49 & 0.11 & 790 & 40 & [13] \\
  RR Lyrae & 24.50 & 0.11 & 790 & 40 & [14] \\
  Tip of the RGB & 24.47 & 0.07 & 785 & 25 & [15] \\
  Eclipsing binary & 24.44 & 0.12 & 770 & 40 & [16] \\
  \hline
   Mean \& std. deviation & 24.39 & 0.08 & 750 & 30 & \\
%754.769+/-26.270 (standard deviation) 24.389+/-0.076
  \hline
 \end{tabular}\\
 \flushleft
 [1]:~\citet{A2004021}; [2]:~\citet{Mould86}; [3]:~\citet{A2004025};
[4]:~\citet{Pritchet87}; [5]:~\citet{Capaccioli89}; [6]:~\citet{A2004026};
[7]:~\citet{Richer90}; [8]:~\citet{Brewer95}; [9]:~\citet{Ostriker97};
[10]:~\citet{Holland98}; [11]:~\citet{Stanek98a}; [12]:~\citet{Durrell01};
[13]:~\citet{A2004006}; [14]:~\citet{Brown04}; [15]:~\citet{A2006007};
[16]:~\citet{P2005002} 
\end{table}

Following the procedure already used in the LMC to obtain precise distance
measurements to several eclipsing binary (EB) systems \citep[see][ and
references therein]{A2003004}, a project was started in 1999 to obtain a direct
distance determination to M31 from EBs \citep{A2003018,P2004001}. The
methodology
involves, at least, two types of observations: photometry, to obtain the light
curve, and spectroscopy, to obtain the radial velocity curve for each component.
By combining the two types of observations the individual masses and radii for
each of the components can be obtained. The remaining needed parameters (luminosities,
temperatures and line-of-sight extinction) can be obtained from the modelling
of the spectral energy distribution \citep[see, e.g.,][]{A2002007} or the
modelling of spectral lines \citep[][ hereafter Paper~I]{P2005002}. The
combination of the analysis results yields an accurate determination of the
distance to the EB system, and hence, to the host galaxy, which is the main
objective of the project. But in addition to providing distance measurements,
the resulting stellar physical properties can be used as powerful diagnostics 
for studying the structure and evolution of stars that have been born in a
chemical environment different from that in the Milky Way. 

The process briefly described above requires high quality light curves to
obtain precise fundamental properties. By the time this project started, the
highest quality EB light curves were those obtained by the DIRECT group
\citep[see][ and references therein]{A2004003}, but the scatter was too
large for a reliable determination of the elements. Therefore, we began a new
photometric survey to obtain high quality EB light curves (see
\S\ref{obs}). With the application of the
difference image analysis algorithm in the data reduction process (see
\S\ref{datared}), precise photometry was obtained not only for the EBs in the
f\mbox{}ield, but also for all the variable stars. 

Two photometric catalogs were compiled (see \S\ref{cats}). The reference
catalog contains photometry, in both $B$ and $V$ passbands, of 236\,238 stars
from which 3\,964 are also in the variable star catalog (containing 437 EBs
and 416 Cepheids). For some of the best quality light curves, corresponding
to the brightest EB systems -- 24 in total, -- a preliminary determination of
the orbital and physical properties is presented (see \S\ref{ebs}). These
systems can be considered potential targets for distance determination and,
indeed, already one of them was used to derive the f\mbox{}irst direct distance
determination to M31 \citepalias{P2005002}.  

\section{Observations}
\label{obs}

The observations were conducted at the 2.5 m Isaac Newton Telescope (INT) in La
Palma (Spain). Twenty one nights in f\mbox{}ive observing seasons were granted between
1999 and 2003 (one season per year). The Wide Field Camera, with four thinned
EEV 4128$\times$2148 CCDs, was used as the detector. The f\mbox{}ield of view of the
WFC at the INT is $33\farcm8\times33\farcm8$, with 0\,$\farcs$33 pixel$^{-1}$
angular scale. With this conf\mbox{}iguration, we found that an exposure time of 15
minutes provides the optimum S/N for stars of $V\sim19$ mag and ensures that no
signif\mbox{}icant luminosity variation occurs during the integration for EBs with a
period longer than one day. 

All observations were centred on the same f\mbox{}ield located at the North-Eastern
part of M31 ($\alpha=00^{\rm h}44^{\rm m}46^{\rm s}$
$\delta=+41\degr38\arcmin20\arcsec$). The f\mbox{}ield of study was selected to
overlap with the DIRECT f\mbox{}ields A, B and C, since the initial main objective of
the project was to obtain high quality photometry of already known EBs. To
reduce blending problems in such a crowded f\mbox{}ield (see Fig. \ref{wfcfov}) all
the images with
PSF FWHM larger than $3\arcsec$ were rejected. As a result, 265 and 259 frames,
with median seeings of 1\,$\farcs$3 and 1\,$\farcs$2, were selected for
further analysis in the $B$ and $V$ passbands, respectively. In addition,
during two nights of the 1999 observing run and one night of the 2001 observing
run, 15 \citet{A2004027} standard stars from 3 f\mbox{}ields (PG0231+051, PG1633+099
and MARK A) were observed. In Table \ref{obsdata}, a complete list of all the
images used in the data reduction process for the f\mbox{}ive observing seasons is
presented. 

\begin{table}[!ht]
 \caption{Number of images used in the data reduction process for each of
the five observing seasons.}
 \label{obsdata}
 \centering
 \begin{tabular}{l r r r r r r}
  \hline
  \hline
  Frame & 1999 & 2000 & 2001 & 2002 & 2003 & Total \\
  \hline
  Granted nights & 3 & 4 & 4 & 5 & 5 & 21 \\
  \hline
  Bias & 5 & 21 & 10 & 46 & 24 & 106 \\
  Darks & 0 & 0 & 0 & 6 & 0 & 6 \\
  $B$ flat-fields & 20 & 31 & 19 & 21 & 42 & 133 \\
  $V$ flat-fields & 23 & 17 & 14 & 27 & 25 & 106 \\
  $B$ Landolt fields & 12 & 0 & 8 & 0 & 0 & 20 \\
  $V$ Landolt fields & 16 & 0 & 11 & 0 & 0 & 27 \\
  $B$ M31 field & 41 & 45 & 45 & 54 & 80 & 265 \\
  $V$ M31 field & 40 & 45 & 43 & 49 & 82 & 259 \\
  \hline
  Total & 157 & 159 & 150 & 203 & 253 & 922 \\
  \hline
 \end{tabular}
\end{table}

\section{Data reduction}
\label{datared}
\subsection{Data calibration}
\label{predia}

The science images for each of the f\mbox{}ive observing seasons were reduced using
the IRAF\footnote{IRAF is distributed by the National Optical Astronomy
Observatories, which are operated by the Association of Universities for
Research in Astronomy, Inc., under cooperative agreement with the NSF.}
package. Only bias, dark and flat-f\mbox{}ield frames corresponding to the same
observing season were used to reduce a given science image. 

Each WFC image was separated into four overscan-subtracted images, one per CCD
frame. After this step, each CCD frame was treated separately during the whole
reduction process. Bad pixels\footnote{Bad pixel positions were obtained from
the webpage: \\ http://www.ast.cam.ac.uk/$\sim$wfcsur/technical/pipeline/} were
also corrected for all images through linear interpolation from the adjacent
pixels. Although this method is not optimal for reducing photometric data, it
yields best results for cleaning the calibration images (bias, dark and
flat-f\mbox{}ield images). 

A master bias was built for each observing season and CCD frame. Each pixel of
the master bias is the median value of the same pixel in all the bias frames.
The master bias was subtracted from the 2002 dark images. The dark images of
the other observing seasons were not used because the dark current was found
to be negligible. The bias-subtracted dark images were combined in the same way
as the bias images (with a median) to produce the 2002 master dark frame. 

A master flat-f\mbox{}ield image was produced for each observing season, CCD frame and
f\mbox{}ilter. The bias and, for the 2002 images, dark frames were subtracted from 
each flat-f\mbox{}ield frame. The resulting frames were corrected for non-linearity
effects\footnote{Coefficients were obtained from the webpage: \\
http://www.ast.cam.ac.uk/$\sim$wfcsur/technical/foibles/} and averaged to
produce the master flat-f\mbox{}ield image. 

The science images were processed by subtracting the bias frame (and also the
dark frame for the 2002 images) and corrected for non-linearity and flat
f\mbox{}ielding. The resulting images are free from most of the instrumental effects,
except for a small area close to the North-Eastern corner of the f\mbox{}ield that is
associated with f\mbox{}ield vignetting (see Fig. \ref{wfcfov}).

\subsection{Difference image analysis}
\label{dia}

\begin{figure}[!ht]
 \resizebox{\hsize}{!}{\includegraphics{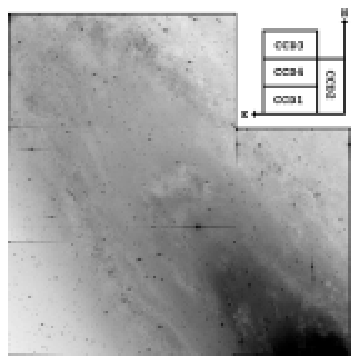}}
 \caption{Reference image of the selected field of view in $V$ filter. The
center of the field is at $\alpha=00^{\rm h}44^{\rm m}46^{\rm s}$ and 
$\delta=+41\degr38\arcmin20\arcsec$}
 \label{wfcfov}
\end{figure}

Given that the f\mbox{}ield under study is highly crowded, a package based on the
image subtraction algorithm \citep{A2002008} was used to obtain the best 
possible photometric precision. This technique has the advantage that variable
stars are automatically detected and that precise photometry can be obtained
even in highly crowded f\mbox{}ields. We have used our own implementation of the
difference image analysis package (DIA) developed by \citet{A2002001}.
The image subtraction algorithm requires a high quality reference image. This
image was created during the f\mbox{}irst step of the process by the DIA for each CCD
and f\mbox{}ilter. The combination of the 15 best seeing images produced two reference
images, one with 0\,$\farcs9$ FWHM for $V$ (see Fig. \ref{wfcfov}) and one with
1\,$\farcs0$ FWHM for $B$. 

The main part of the process is the determination of the convolution kernels.
The kernels are constructed to match the reference image with each one of the
remaining frames under study (supposedly with different seeing values). The DIA
kernels are composed of Gaussian functions with constant widths and modif\mbox{}ied by
polynomials. In our case, the kernels consist
of three Gaussians with polynomials of orders 4, 3 and 2. A space-varying
polynomial of order 4 has also been used to account for the spatial variation
across the image. The resulting kernels are convolved with the reference image
and subtracted from the original frames to obtain images containing only
objects with brightness variations (plus noise). 

The variability image is constructed from the differentiated images.  This
variability image contains all pixels considered to have light level 
variations. A pixel is considered to vary if it satisf\mbox{}ies at least one of
the following two conditions: 
\begin{itemize}
 \item There are more than 4 consecutive observations deviating more than
3$\sigma$ (where $\sigma$ is the standard deviation) from the base line flux in
the same direction (brighter or fainter). 
 \item There are more than 10 overall observations deviating more than
4$\sigma$ from the base line flux in the same direction. 
\end{itemize}

The variability image is used to identify the variable stars. The reference
image PSF is correlated with the variability image and all local maxima
with correlation coefficients in excess of 0.7 are considered to be variable
stars. Once the variable stars are located, PSF photometry is performed on all
the differentiated images to obtain the differential fluxes. The differential
fluxes can be transformed to the usual instrumental magnitudes through the
following expression:
\begin{equation}
 \label{fins}
 m_i=C-2.5\log(f_0+a\Delta f_i)
\end{equation}
\noindent where $\Delta f_i$ is the differential flux of a star in the $i$-th
observation, $f_0$ is the base line flux on the reference image, $C$ is a
zeropoint constant and $m_i$ is the instrumental magnitude of a star in the
$i$-th observation. 

The photometric package of DIA does not properly account for contamination from
nearby stars in crowded f\mbox{}ields (such as our case). Therefore, the $f_0$ values
obtained with the DIA can potentially be severely overestimated. To obtain
precise values for $f_0$, DAOPHOT PSF photometry \citep{A2004023} was applied
to the reference image. Given that DIA and DAOPHOT photometry use different PSF
def\mbox{}initions and functional forms, a scaling factor ($a$) is needed to transform
the DIA fluxes into the DAOPHOT flux units \citep[see ][ for an extensive
discussion]{A2003001}. A synthetic image was created to compute the scaling
factors. The synthetic image contains the representation of the DAOPHOT PSFs at
the position of each variable star. The scaling factor at the position of each
variable star was obtained from the comparison with the DIA photometry of each
PSF. For similar PSFs, the scaling factor can be obtained with an error of
0.3\% or less. 

This process provided instrumental magnitudes for all the variable stars as
well as all the stars in the reference image. Two main sources of potential
systematic errors exist with this procedure, $f_0$ and $a$, but they affect the
amplitudes of the light curves. However, as it is shown below in \S\ref{ebs},
the amplitudes of the f\mbox{}itted EBs are perfectly compatible with other results
given in the literature. 

\subsection{Standard photometry}
\label{stdphot}

To transform instrumental to standard magnitudes, the observed \citet{A2004027}
standard stars were used. Although the standard stars were observed at
different times during the night, the number of collected frames is not enough
to accurately determine the atmospheric extinction. Therefore, as a f\mbox{}irst step,
the M31 images obtained during the same night as the standard stars (30 images
in total) were used to compute the coefficients needed to account for the
atmospheric extinction. Aperture photometry of 20 bright, but not saturated,
and isolated stars on the M31 frames was performed. For each night and f\mbox{}ilter,
a linear f\mbox{}it, with dispersions ranging from 0.008 mag to 0.03 mag, was
obtained. 

As a second step, aperture photometry was performed on the standard stars.  The
instrumental magnitudes obtained were corrected with the atmospheric extinction
coefficients. The resulting values were used to f\mbox{}ind the transformation
coefficients between the instrumental magnitudes and the standard magnitudes.
For each night and CCD the following relationship was determined:
\begin{eqnarray}
\label{mins}
 v-V=A_1+A_2(B-V) \\
 b-v=A_3+A_4(B-V)
\end{eqnarray}
\noindent where $b$ and $v$ are the instrumental magnitudes, $B$ and
$V$ are the standard magnitudes and $A_i$ are the transformation coefficients.
The resulting mean standard deviation of the f\mbox{}its is 0.02 mag. 

The transformation coefficients cannot be directly applied to the M31
instrumental magnitudes because the latter are based on the reference image,
which is a combination of different images. In addition, the transformation
coefficients are determined from aperture photometry, while the M31
instrumental magnitudes have been determined from PSF photometry. For this
reason, the only way to obtain standard magnitudes for the M31 stars is to
apply the transformation coefficients to the frames obtained during the same
nights as the standard stars. 

PSF photometry is needed to f\mbox{}ind precise standard magnitudes of a reasonable
number of objects in the M31 f\mbox{}ield since aperture photometry can only be
applied to isolated stars. A sample of bright and isolated stars was employed
to determine a scaling value used to transform the PSF magnitudes into aperture
photometry values for each one of the 30 M31 frames. The scaling values and the
transformation coefficients were used to obtain standard magnitudes for 18\,426
objects in the f\mbox{}ield. A systematic difference was observed for the standard
magnitudes obtained from the 2001 frames and, therefore, the obtained values
were rejected. The standard magnitudes of every star in all the remaining
frames (22 in total) were averaged and the standard deviation was considered a
good estimation of its uncertainty. 

\begin{figure}[!ht]
 \resizebox{\hsize}{!}{\includegraphics{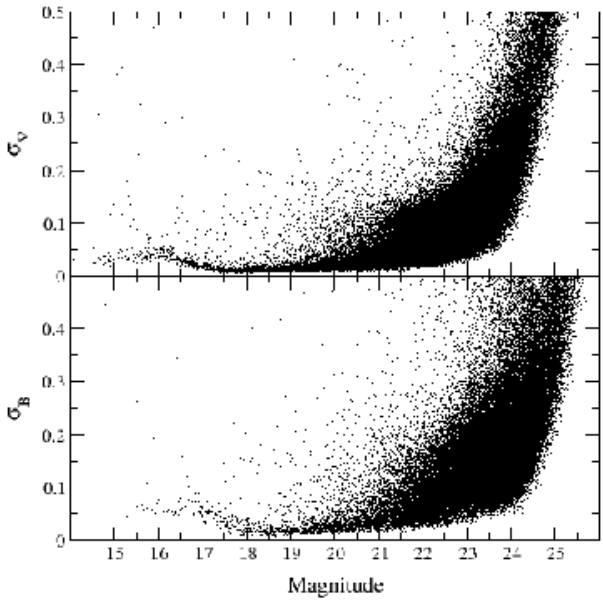}}
 \caption{Error distribution for the entire photometric catalog containing
236\,238 objects with photometry in both $V$ and $B$ passbands.}
 \label{magerr}
\end{figure}

The standard magnitudes resulting from this process were used to determine new
transformation coefficients ($A_i$) for the reference images. Only 534
non-variable and non-saturated objects detected in all the frames and with an
error below 0.04 mag were used for this purpose. This sample has good color
and magnitude coverage, with $-0.3<B-V<1.7$ and $17.5<V<21.0$, providing f\mbox{}its
with dispersions ranging from 0.013 mag to 0.019 mag, depending on the CCD.
The obtained coefficients were applied to the reference images. The objects
detected in the $V$ frame have been cross-matched with the objects detected in
the $B$ frame. Only the objects identif\mbox{}ied within 0\,$\farcs$33 (one pixel)
have been included in the reference catalog, providing standard photometry for
236\,238 objects. The error distributions, shown in Fig \ref{magerr}, have
37\,241 objects with errors below 0.1 mag in both passbands. The limiting
magnitudes of the photometric catalog are around $V\simeq25.5$ and
$B\simeq26.0$, and it is estimated to be complete up to $V\simeq22.3$ mag and
$B\simeq23.5$ mag. The same process was applied to transform the $m_i$ values
in Eq. \ref{fins} to standard photometry in both $B$ and $V$ passbands for a
total of 3\,964 variable stars. 

\subsection{Astrometry}

Standard coordinates for all detected objects were computed from several
reference stars in the GSC 2.2.1\footnote{Data obtained from:
http://cdsweb.u-strasbg.fr/viz-bin/Vizier}. To ensure that the reference stars
are uniformly distributed, each CCD was divided into 3$\times$6 sectors.  Three
reference stars (with $V\simeq18$) were identif\mbox{}ied manually in each
sector, yielding an initial list of 54 reference stars per CCD. Fits using
third order linear equations with 2$\sigma$ scatter clipping were applied to
determine the transformation of coordinates. After the iterative process at
least two reference stars had to remain in each sector. Otherwise, an
additional reference star was selected in the corresponding sector and the
entire process was repeated. 

\begin{figure}[!ht]
 \resizebox{\hsize}{!}{\includegraphics{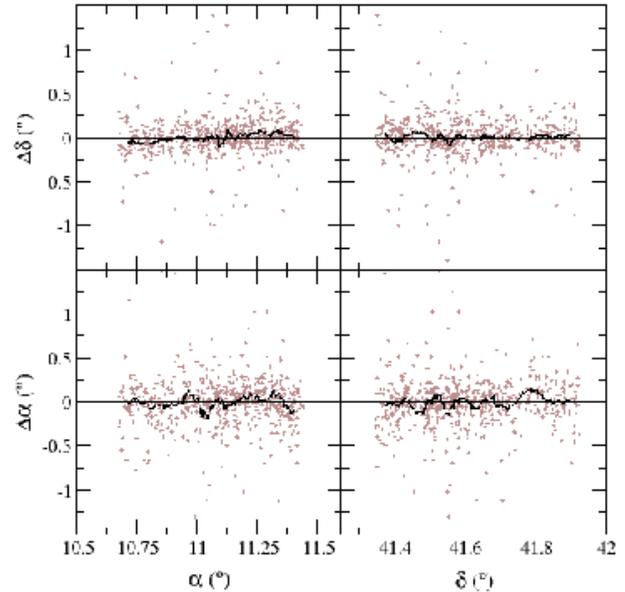}}
 \caption{Differences between the coordinates obtained from our procedure and
the coordinates from the GSC 2.2.1 catalog for 724 identified stars (circles).
The line represents a running average.}
 \label{coordszp}
\end{figure}

The resulting coordinates have been compared with those in the GSC 2.2.1
catalog. A total of 724 objects with a position difference of less than
$3\arcsec$ in the two catalogs were identif\mbox{}ied. As can be seen in Fig.
\ref{coordszp}, no systematics trends appear in the comparisons, which have
dispersions of $\sigma_\alpha=0$\,$\farcs$16 and $\sigma_\delta=0$\,$\farcs$12.

\subsection{Periodicity search}

Since the observations were performed in f\mbox{}ive annual observing seasons, the
data is very unevenly sampled. To overcome this major drawback for the
determination of periodicities, a program based on the analysis of variance
\citep{A2003002} was used to compute the periodograms for each variable star
and f\mbox{}ilter ($B$ and $V$). The periodograms were computed with 7 harmonics
between 1 and 100 days because this setup was found to provide optimal results
for the detection of EB systems, which constitute the main objective of the
present survey. The lower limit in the period search is set from the
consideration that EBs with periods below 1 day are likely to be foreground
contact systems and therefore not of the interest of the survey. In addition,
given the relatively long exposure time for each observation in the present
survey (15 minutes), important brightness variations during the integration
time could occur for the short period variables. These factors make the
computational effort of extending the period search below 1 day not worthwhile.
On the opposite end, the upper limit of 100 days was selected to detect as many
long period Cepheids as possible.

For each variable star, two periodograms (one in $B$ and one in $V$) were used
to obtain a consistent period determination. The resulting light curves were
visually inspected to locate well-def\mbox{}ined variability patterns, thus leading to
the identif\mbox{}ication of a total of 437 EBs and 416 Cepheids. For several EBs and
Cepheids, the period determination was a multiple of the true period and,
therefore, it had to be recomputed. 

For each variable star, the time of minimum of both light curves ($B$ and $V$)
was used to compute a reference time. The time of minimum was computed from the
Fourier series \citep[$F^{\rm (7)}(t)$ in][]{A2003002} that best f\mbox{}it each light
curve. 

\section{The catalogs}
\label{cats}
\subsection{Reference catalog}
\label{refcat}

Photometry of the 236\,238 objects detected in the reference images was grouped
into the reference catalog. The reference catalog (Table 3) contains the object
identif\mbox{}ier, the right ascension, the declination, the $V$ standard
magnitude in the reference image, the $V$ standard error, the $B$ standard
magnitude in the reference image, the $B$ standard error and, in case that
there is a previous identif\mbox{}ication, the corresponding identif\mbox{}ier.
The catalogs cross-matched with our reference catalog are GSC 2.2.1, DIRECT,
LGGS \citep{A2006003} and \citet[][ hereafter \citetalias{A2005015}]{A2005015}.
Photographic catalogs are identif\mbox{}ied in the previous works, specially in
LGGS, and they are not cross-identif\mbox{}ied here. 

According to the IAU recommendations, the identif\mbox{}ier was built from the
object position and taking into account the observational angular resolution.
The resulting format is M31\,JHHMMSSss+DDMMSSs. For the variable stars, the
acronym was changed to M31V to indicate that they are also in the variable star
catalog (see \S\ref{varcat}). 

\begin{figure}[!ht]
 \resizebox{\hsize}{!}{\includegraphics{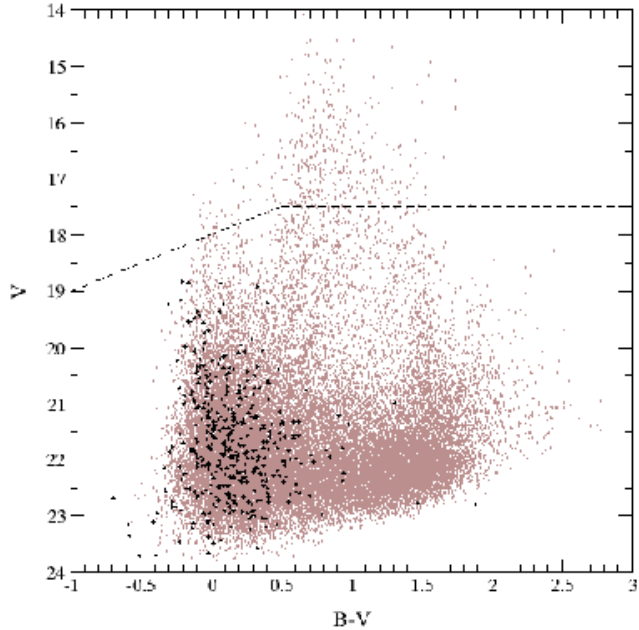}}
 \caption{Color--magnitude diagram for 37\,241 objects with photometric errors
lower than 0.1 mag in both $B$ and $V$ in the reference catalog (grey dots) and
437 EBs (black circles). Stars above the dashed lines are saturated.}
 \label{vbvebs}
\end{figure}

To show the general photometric properties of the obtained catalog, the
color--magnitude diagram for the 37\,241 objects with a photometric error below
0.1 mag in both $B$ and $V$ is shown in Fig. \ref{vbvebs}. From this diagram it
can be seen that most of the stars in this catalog are stars at the top end of
the main sequence, with some foreground giants and stars at the tip of the red
giant branch. 

\begin{figure}[!ht]
 \resizebox{\hsize}{!}{\includegraphics{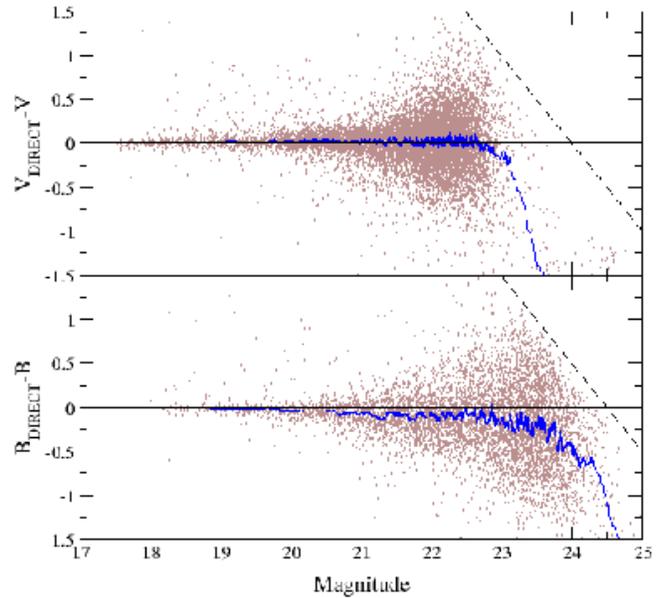}}
 \caption{Differences with DIRECT standard magnitudes versus the magnitudes in
the reference catalog. The comparison has 14\,717 stars in $V$ and 7\,499 in
$B$. The dashed lines show the limiting magnitudes.}
 \label{directzp}
\end{figure}

\begin{figure}[!ht]
 \resizebox{\hsize}{!}{\includegraphics{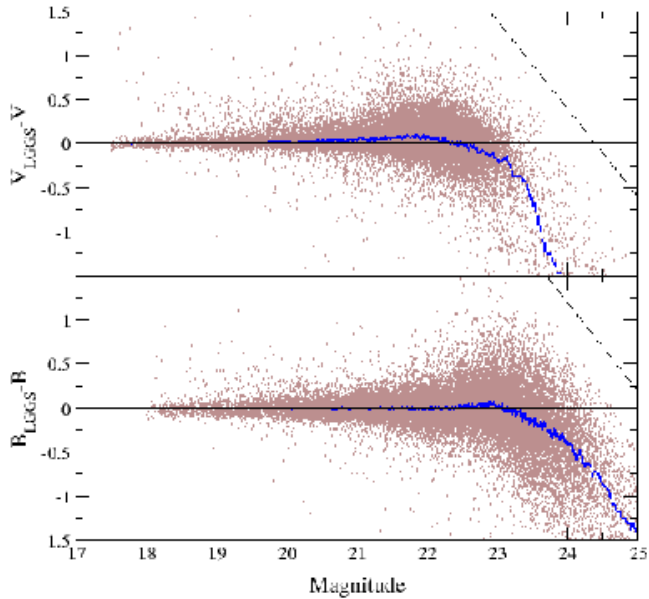}}
 \caption{Differences with LGGS standard magnitudes versus the magnitudes in
the reference catalog for a total amount of 36\,353 stars in each passband. The
dashed lines show the limiting magnitudes.}
 \label{lggszp}
\end{figure}

The obtained standard photometry depends on a number of zeropoint calibrations
(\S\S \ref{dia} and \ref{stdphot}). Taking into account that an error in the
standard magnitudes has a direct impact on the distance determination, it is
extremely important to ensure that no systematic errors affect the photometry.
For this reason, it is worthwhile to check the consistency of the standard
magnitudes given here. Fortunately, there are two CCD surveys with $B$ and $V$
photometry of a large number of stars that overlap with the f\mbox{}ield under study:
DIRECT \citep{A2004003} and LGGS \citep{A2006003}. The non-saturated objects of
our survey ($V>17.5$ and $B>18.0$) were cross-identif\mbox{}ied with the magnitudes
reported by the DIRECT group\footnote{Data source:
ftp://cfa-ftp.harvard.edu/pub/kstanek/DIRECT/} and by the LGGS
group\footnote{Data source: http://www.lowell.edu/users/massey/lgsurvey/}. For
the DIRECT survey, a total of 14\,717 and 7\,499 objects were identif\mbox{}ied in $V$
and $B$ passbands, respectively (see Fig. \ref{directzp}). The comparison with
the LGGS survey provided 36\,353 common objects (see Fig. \ref{lggszp}). 

Although some trend is observed for the DIRECT $B$ magnitudes, the LGGS $B$
magnitudes are completely consistent with the magnitudes in our survey.
Therefore, the $B$ standard magnitudes in the present paper are expected to
have no systematic error. Regarding the $V$ magnitudes, some low-level
systematics have been observed with the two catalogs. The DIRECT $V$ magnitudes
are larger by about 0.02 mag than the values in the reference catalog, and the
LGGS comparison shows an increasing trend towards larger magnitudes. The 5\,768
objects with $V$ magnitudes in the three catalogs were used to study the
observed systematics. The comparisons between the three catalogs always reveal
some kind of trend and, therefore, it was impossible to conclude weather there
are any (low-level) systematics affecting the $V$ standard magnitudes. In any
case, the differences for the brightest stars ($V<20.5$ mag) are well below
0.03 mag. Taking into account that all the EBs than can be used for distance
determination have $V<20.5$ (\S \ref{lcfit}), this is the maximum systematic
error that can exist in their $V$ standard magnitudes. 

\subsection{Variable star catalog}
\label{varcat}

The 3\,964 variable stars identif\mbox{}ied in the reduction process were grouped in
the variable star catalog (Table 4). For each variable star, the corresponding
identif\mbox{}ier (see \S\ref{refcat}), the right ascension, the declination, the
intensity--averaged $V$ magnitude \citep[Eq.  $(9)$ in][]{A2005001}, the mean
$V$ error, the intensity--averaged $B$ magnitude, the mean $B$ error, the
number of observations in $V$, the number of observations in $B$, the reference
time (in HJD), the period (in days) and a label, indicating weather the
variable star was identif\mbox{}ied as an EB or a Cepheid, are provided. 

\begin{figure}[!ht]
 \resizebox{\hsize}{!}{\includegraphics{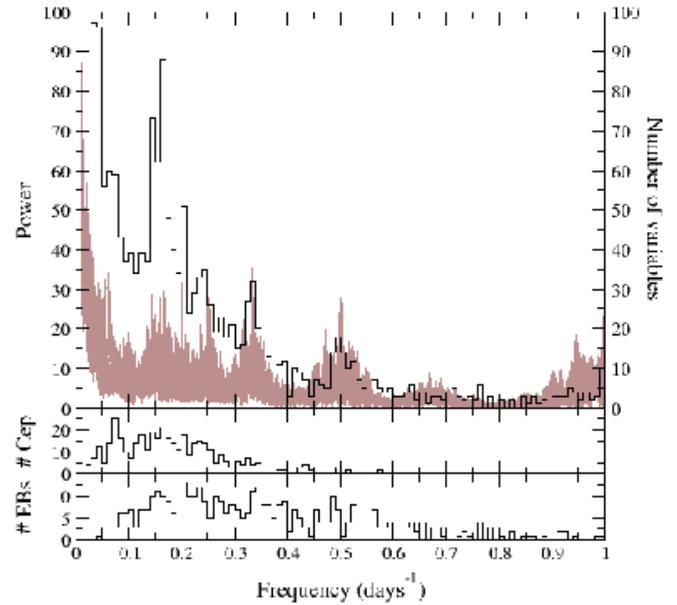}}
 \caption{Top: Window function for the studied period range (lines) and
frequency distribution for all the stars in the variable star catalog
(histogram). Middle: Frequency distribution for the Cepheids in the variable
star catalog. Bottom: Frequency distribution for the eclipsing binaries in the
variable star catalog.}
 \label{winfunc}
\end{figure}

Although a period estimate is given for each variable star, obviously not all
the variable stars are periodic. When comparing the window function with the
period distribution for all variable stars (see Fig. \ref{winfunc}), it is
clear that many of the period determinations are just an alias introduced by
the window function, specially for those over f\mbox{}ive days. Therefore, a
signif\mbox{}icant number of the variable stars are, in fact, non-periodic variables
or variables with a period out of the studied range (1--100 days). All light
curves (time series photometry) are also provided (Table 5) together with the
variable star catalog. For each variable star and observation, the observed
time (in JD), the standard magnitude ($V$ or $B$) and the standard error are
given. 

\section{Eclipsing binaries}
\label{ebs}
\subsection{Complete sample}

The color--magnitude diagram (see Fig. \ref{vbvebs}) reveals that most of the
detected EB systems contain high-mass components belonging to the top of the
main sequence. Taking into account that most of the systems have periods
shorter than 10 days (and all of them have periods shorter than 30 days), a
large fraction of EB systems are non-detached systems.

\begin{figure}[!ht]
 \resizebox{\hsize}{!}{\includegraphics{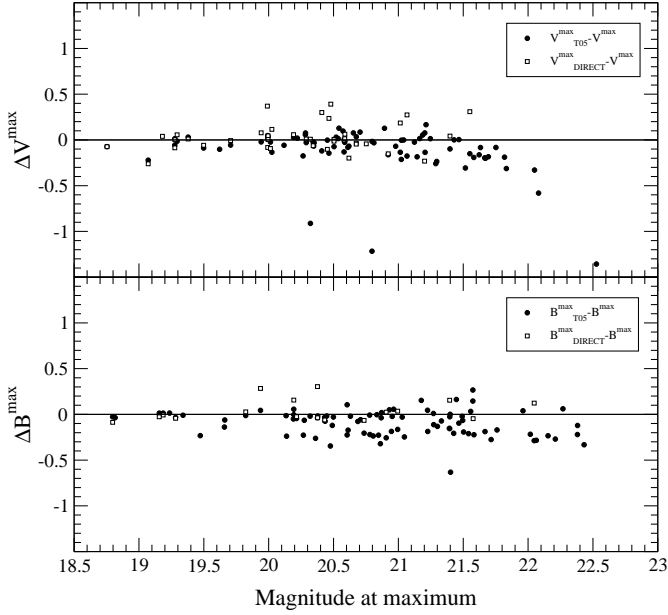}}
 \caption{Differences between the DIRECT maximum magnitudes (open squares), the \citetalias{A2005015} maximum magnitudes (filled circles) and the maximum
magnitudes for all the crossed EBs of the variable star catalog.} 
 \label{ebszp}
\end{figure}

Part of the images in the present survey were also reduced by
\citetalias{A2005015} and reported the detection of 127 EB systems. Of those,
123 stars have been detected in our reference catalog, 92 are in the variable
star catalog and 90 have also been classif\mbox{}ied as EBs. The maximum
$V^{\rm max}$ and $B^{\rm max}$ values for the 90 systems identif\mbox{}ied
were matched with the magnitudes presented in \citetalias{A2005015} (see Fig.
\ref{ebszp}). A clear systematic trend can be observed in the $V$
f\mbox{}ilter. In addition, Fig.  \ref{ebszp} also reveals that the $B^{\rm
max}_{\textrm{\scriptsize\citetalias{A2005015}}}$ magnitudes are lower than our
$B^{\rm max}$ values.

When inspecting the origin of such discrepancy, we observed that \emph{all} the
EBs labeled f2BEB and f3BEB in \citetalias{A2005015} had lower magnitude values
than the EBs in our variable star catalog. Since 14 of these EBs have also
DIRECT $V^{\rm max}$ magnitudes, and 5 of them have DIRECT $B^{\rm max}$ values
\citep{Kaluzny98, Stanek98b, Stanek99, A2003006}, the mean differences were
computed (Table \ref{diffebs}). Although the number of cross-matched EBs is
small, their mean differences seem to indicate that the magnitudes reported in
\citetalias{A2005015} for the f2BEB and f3BEB EBs suffer from a systematic
offset. On the other hand, the magnitude values for the remaining EBs (f1BEB
and f4BEB) are fairly compatible in the three catalogs. 

\begin{table}[!ht]
 \renewcommand{\thetable}{6}
 \caption{Mean differences for the magnitudes at maximum of the EBs labeled
f2BEB and f3BEB in \citetalias{A2005015}.}
 \label{diffebs}
 \centering
 \begin{tabular}{l r@{$\pm$}l}
  \hline
  \hline
   Comparison & \multicolumn{2}{ c }{Differences} \\
  \hline
   $V^{\rm max}_{\textrm{\scriptsize DIRECT}}-V^{\rm max}$ & 0.062 & 0.047 \\ 
   $V^{\rm max}_{\textrm{\scriptsize\citetalias{A2005015}}}-V^{\rm max}$ & -0.104 & 0.015 \\
   $V^{\rm max}_{\textrm{\scriptsize\citetalias{A2005015}}}-V^{\rm max}_{\textrm{\scriptsize DIRECT}}$ & -0.166 & 0.057 \\ 
%$I-D=-0.062020\pm0.046691341$, $T-I=-0.104409\pm0.014752286$ and
%$T-D=-0.166429\pm0.056593904$. 
  \hline
   $B^{\rm max}_{\textrm{\scriptsize DIRECT}}-B^{\rm max}$ & 0.054 & 0.039 \\
   $B^{\rm max}_{\textrm{\scriptsize\citetalias{A2005015}}}-B^{\rm max}$ & -0.214 & 0.026 \\ 
   $B^{\rm max}_{\textrm{\scriptsize\citetalias{A2005015}}}-B^{\rm max}_{\textrm{\scriptsize DIRECT}}$ & -0.268 & 0.046 \\
%$D-I=0.054130\pm0.03901402$, $T-I=-0.213870\pm0.025587773$ and
%$T-D=-0.268000\pm0.04641138$
  \hline
 \end{tabular}
\end{table}

\subsection{Light curve fitting}
\label{lcfit}

From all the variable stars identif\mbox{}ied as EBs only those systems with a precise
determination of their fundamental properties can be used as distance
determination targets. Taking into account that medium--resolution spectra are
needed to obtain precise fundamental properties, and considering the fact that
the largest currently available facilities are the 8--10~m class telescopes,
only the brightest EB systems (with $V^{\rm mean}<20.5$) were selected for
further study. In addition, the most precise fundamental properties can be
achieved only for those systems with deep eclipses, and thus we further
selected only those EBs with $\Delta B\geq 0.2$ mag and $\Delta V\geq 0.2$ mag.

The criteria above provided a list of 29 systems from the initial sample of 437
EBs. Of these, 5 systems were rejected because of the large scatter in their
light curves (probably from the contamination of a brighter nearby star),
leaving a total amount of 24 EBs selected for detailed further analysis. The
particular properties of each EB can have an important influence on the precise
determination of its fundamental parameters. To obtain the best distance
determination targets, a preliminary f\mbox{}it was performed, with the 2003 version
of the \citet{A2005013} program (W-D), to the selected sample of 24 EBs.  

The f\mbox{}itting process was carried out, in an iterative manner, by considering
both light curves ($B$ and $V$) simultaneously. For those systems previously
identif\mbox{}ied by the DIRECT group their $V$ light curve was also included in the
f\mbox{}itting process. This provided an additional check on the consistency of the
resulting photometry, because the DIRECT photometry was calculated from a PSF
rather than using a DIA algorithm. Once a converging solution was achieved, a
$4\sigma$ clipping was performed on all the light curves to eliminate (normally
a few) outlier observations. 

A certain conf\mbox{}iguration has to be assumed for each solution of W-D. Two basic
conf\mbox{}igurations were considered for each EB system: detached and semi-detached.
Generally, the f\mbox{}inal f\mbox{}it in each one of the conf\mbox{}igurations provided some clues
about the real conf\mbox{}iguration of the EB system. However, all systems showing
non-zero eccentricity (secondary eclipse phase not exactly at 0.5) were
considered to be detached EB systems. In some specif\mbox{}ic cases, a sinusoidal
$O-C$ was observed when studying the f\mbox{}it residuals, with one quadrature being
brighter than the other. This is known as the O'Connell effect. In the case of
interacting high-mass systems, such effect can be explained by the presence of
an equatorial hot spot on the surface of one the components. Such hot spot is
supposed to the consequence of impacting material arising from mass transfer
between the components in a Roche-lobe f\mbox{}illing conf\mbox{}iguration (semi-detached).
The exact parameters of the spot cannot be obtained from the available data
\citep[strong degeneracies exist;][]{A2003004} and, therefore, the spot
parameters presented here are not necessarily physically valid but only capable
of providing a good f\mbox{}it to the variations in the light curves.

Each EB is a particular case and careful selection of the adjustable parameters
was performed individually. In general, the parameters f\mbox{}itted with W-D are: the
time of minimum ($t_{\rm min}$), the period ($P$), the inclination ($i$), the
effective temperature of the secondary ($T^{\rm eff}_2$), the normalized
potential of primary ($\Omega_1$), the normalized potential of secondary
($\Omega_2$), for detached systems only, and the luminosity of the primary
($L_1$). For the eccentric systems, in addition, the eccentricity ($e$) and the
argument of the periastron ($\omega$) were also f\mbox{}itted. In some cases of
semi-detached systems, the mass ratio ($q$) was treated as a free parameter.
Finally, a third light ($l_3$) contribution was included for the f\mbox{}inal
solutions of the EB systems. Taking into account that the f\mbox{}ield under study is
extremely crowded, it is not surprising that some EBs may have a signif\mbox{}icant
third light contribution (see \S\ref{blend} for a more extended discussion on
this topic). In addition, the obtained values of the third light contribution
provide a f\mbox{}irst check on the realistic determination of the fundamental
properties and give an indication that the scaling factor determinations ($a$
in Eq. \ref{mins}) are correct. Indeed, an error in the scaling factor has
exactly the same effect as a third light contribution. The limb-darkening
coefficients (square-root law) were computed at each iteration from Kurucz
ATLAS9 atmosphere models and the gravity brightening coefficients, as well as
the bolometric albedos, were f\mbox{}ixed to unity (assuming components with radiative
envelopes). 

\begin{figure*}[!ht]
 \resizebox{\hsize}{!}{\includegraphics{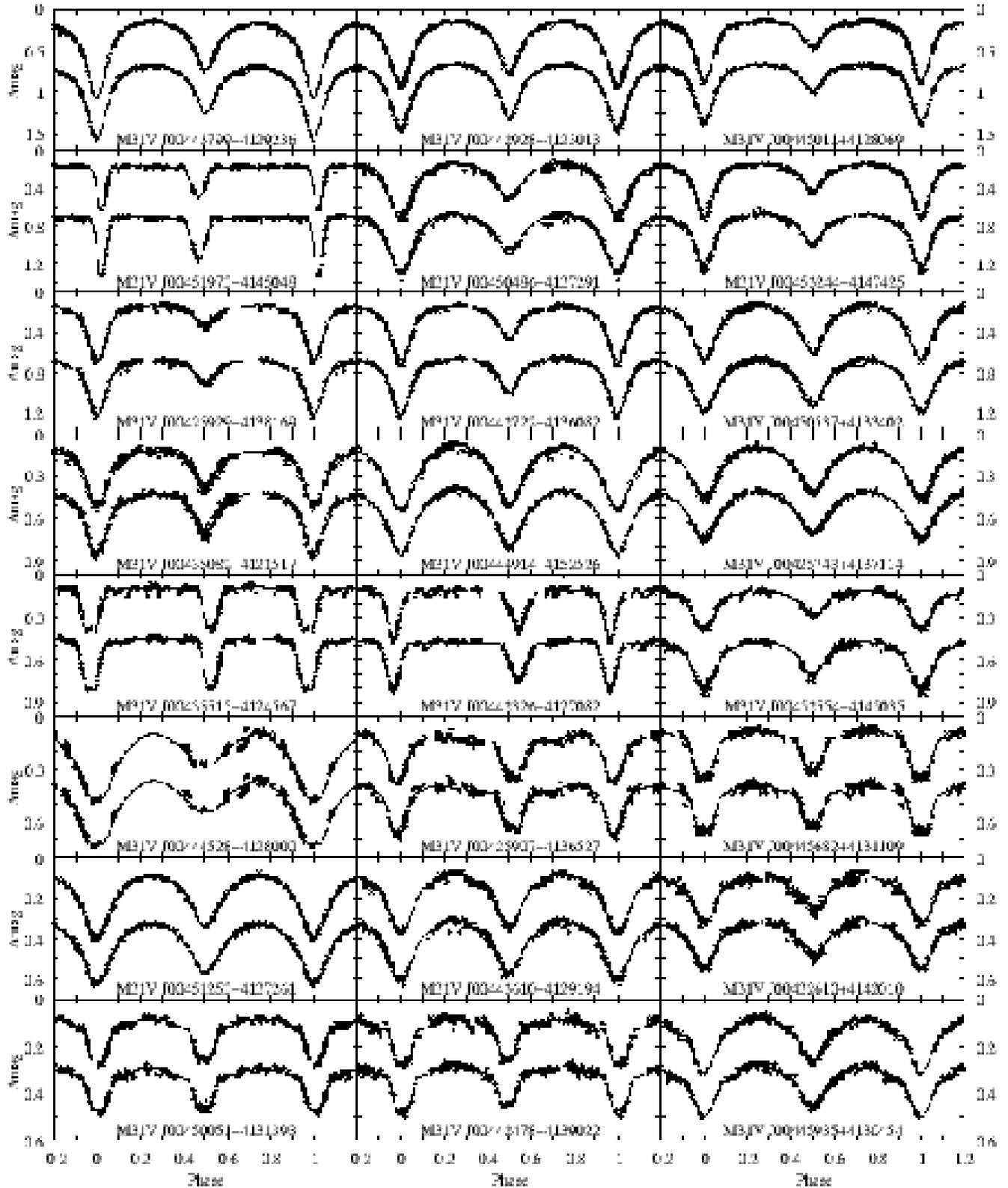}}
 \caption{$B$ (up) and $V$ (down) light curves for the 24 fitted EBs. The
corresponding Wilson--Devinney fits are also plotted (lines).}
 \label{fits}
\end{figure*}

Note that most of the solutions have a large number of free parameters. To some
extent, the values presented here rely on the adoption of the mass ratio and,
therefore, the f\mbox{}it results (in Table \ref{fitstab} and Fig. \ref{fits}) should
be regarded as preliminary until radial velocity curves are obtained. For some
of the binaries the values given in Table \ref{fitstab} need additional
comments:
\begin{itemize}
 \item M31V\,J00444528+4128000. This is the brightest EB system detected.  A
converging f\mbox{}it could not be achieved because of variations of the light curve
at the second quadrature. The reason is so far unidentif\mbox{}ied but it could be due
to a varying hot spot, an accretion disk or intrinsic variability of a
supergiant component. This makes the system an interesting target for future
follow-up observations. 
 \item M31V\,J00442326+4127082. The DIRECT light curve reveals a variation in
the argument of periastron thus indicating the presence of rapid apsidal motion
(1.9$\pm$0.6 deg~year$^{-1}$). Even taking into account the important third
light contribution, the radial velocity curves can provide valuable information
to study the real nature of this detached, eccentric system. 
 \item M31V\,J00443799+4129236. Radial velocity curves have already been
acquired with the 8-m Gemini North telescope and more accurate fundamental
parameters have been derived, in \citetalias{P2005002}, from the f\mbox{}it of both
radial velocity and light curves, as well as from the f\mbox{}it of the atmospheric
parameters from the spectra. However, we decided to consider this system within
the current homogeneous analysis and thus the comparison of the parameters
presented here and those in \citetalias{P2005002} provides an idea of the
accuracy of the parameters in semi-detached systems when the mass ratio is
unknown. 
 \item M31V\,J00451973+4145048. A likely large error in the scaling factor 
causes a large negative $V$ third light contribution. In any case, a correct
third light contribution is obtained in the DIRECT light curve and, therefore,
the fundamental parameters are expected to be correctly determined.  
 \item M31V\,J00444914+4152526. An equatorial hot spot was assumed for this EB
system. The best f\mbox{}it yields a spot with a temperature ratio of 1.2, at a
longitude of $246\pm14$ deg and a size of $17\pm3$ deg. 
 \item M31V\,J00442478+4139022. The large scatter of the DIRECT $V$ light
curve prevented a reliable f\mbox{}it from being achieved. Consequently, the DIRECT
light curve was not used.
 \item M31V\,J00445935+4130454. A very slight O'Connell effect was observed in
this EB system, suggesting the presence of a hot spot. The best f\mbox{}it, with a
temperature ratio of 1.1, yields a spot at $280\pm60$ deg and a size of
$21\pm11$ deg. 
\end{itemize}

The obtained values provide a good estimation of the potential of each EB
system to become a good distance determination target. In general, the best
distance determination targets should have low third light contribution and a
luminosity ratio close to unity. On one hand, an important third light
contribution would lead to an underestimation of the observed standard
magnitudes for the EB system. Although the third light can be modelled, the
resulting values may have uncertainties that are large enough to have a
sizeable negative effect on the accuracy of the distance determination. On the
other hand, EBs with luminosity ratios close to unity will have spectra with
visible lines for the two components, thus making possible the determination of
accurate fundamental properties. Taking into account the obtained values,
several EBs have already been selected as targets to obtain radial velocity
measures and, in fact, already one of them has been used to determine the f\mbox{}irst
direct distance to M31 \citepalias{P2005002}.

\subsection{Blending}
\label{blend}

Special care has to be taken with the magnitudes reported in the present paper.
As mentioned above, the FWHM of the reference images is around $1\arcsec$. At
the distance of M31 \citepalias[$\sim$770 kpc,][]{P2005002} this corresponds to
more than 3.5 pc. Therefore, it is highly probable that a large fraction of the
identif\mbox{}ied objects in the reference catalog are, in fact, a blend of several
stars. In addition, the blending effect can signif\mbox{}icantly modify the observed
amplitudes of variable stars and therefore their derived mean magnitudes can
also be in error. \citet{A2003009} studied this problem for Cepheid variables
in M31 and arrived at the conclusion that blending affect the $V$ Cepheid
flux at the 19\% level.

It has been noted before \citepalias{A2005015} that such blending would
seriously hamper the accurate determination of fundamental properties and
distances of M31 EBs. While this is true for Cepheid variables, EB systems are
a particular case of variable stars in which their variations obey a
well-def\mbox{}ined physical model. Therefore, blending effects are easily and
seamlessly incorporated in the EB modelling as an additional (third) light
contribution. While special care has to be taken (e.g., check for physical
consistency between the third light values for different passbands), the
presence of a third light does not preclude the determination of reliable
properties for the components. In any case, no systematic errors in the
distances of EBs will arise from blending because the analysis can detect its
effects in a natural manner. Note that the mean $V$ third light value for the
24 f\mbox{}itted systems in our analysis is of 10\% and it would be around 20\% from a
rough scaling to single stars. This is in excellent agreement with the value
derived by \citet{A2003009}.
%l_3(B,22)=0.109849+/-0.013476636
%l_3(V,22)=0.107612+/-0.016673149
%l_3^{DIRECT}(V,15)=0.111251+/-0.021533529

\section{Conclusions}

Deep and high-quality time-series photometry has been obtained for a region in
the North--Eastern quadrant of M31. The present survey is the deepest
photometric survey so far obtained in M31. The photometry has been checked for
compatibility with that from other surveys and, therefore, our resulting
catalog can be used as a reference to extend the time baseline of future
stellar surveys in M31.

The study of the variable star population has revealed about 4\,000 variables.
Further analysis is needed for most of them for a proper classif\mbox{}ication but
over 800 variable stars have already been identif\mbox{}ied as EBs and Cepheids. The
study of the Cepheid variables will be the subject of a forthcoming publication
\citep{P2006002}. In the current paper we have presented the analysis of the EB
population, which has resulted in an extensive list of EBs suitable for
accurate determinations of the components' physical properties and their
distances. The catalog provided here constitutes an excellent masterlist to
select systems for further analysis. For example, the resulting physical
properties will be an important tool to study stellar evolution of massive
stars in another galaxy and, therefore, in a completely independent chemical
environment. But, in accordance with the main goal of the the present survey,
the technique of determining accurate distances from EBs has been demonstrated
in \citetalias{P2005002} to harbor great potential. Full analysis of an
additional 5--10 EBs from the sample provided here can result in a distance
determination to M31 with a relative uncertainty of 2--3\% and free from most
systematic errors. This will represent the most accurate and reliable distance
determination to this important Local Group galaxy.

\begin{acknowledgements}
The authors are very grateful to \'A. Gim\'enez for encouragement and support
during the early stages of the project and for his participation in the initial
observing runs. Thanks are also due to P. Wo\'zniak for making their DIA code
available to us, to A. Schwarzenberg-Czerny for his useful comments on the
Analysis of Variance technique and to the LGGS and DIRECT teams for making all
their data publicly available in FTP form. This program has been supported by
the Spanish MCyT grant AyA2003-07736. F.~V. acknowledges support from the
Universitat de Barcelona through a BRD fellowship. I.~R. acknowledges support
from the Spanish MEC through a Ram\'on y Cajal fellowship.
\end{acknowledgements}

\bibliographystyle{aa}
\bibliography{5667}

\begin{table*}[!t]
\renewcommand{\thetable}{7}
 \caption{Fundamental properties of the 24 EBs with Wilson--Devinney fits (see
text for nomenclature).}
 \label{fitstab}
 \centering
 {\scriptsize
 \begin{tabular}{l l l r@{$\pm$}l r@{$\pm$}l r@{$\pm$}l r@{$\pm$}r r@{$\pm$}l r@{$\pm$}l}
  \hline
  \hline
  Identif\mbox{}ier   & Conf.$^a$ & $V^{\rm max}$ & \multicolumn{2}{ c }{P}      & \multicolumn{2}{ c }{$i$}   & \multicolumn{2}{ c }{$e$} & \multicolumn{2}{ c }{$\omega$} & \multicolumn{2}{ c }{$R_1/a^b$} & \multicolumn{2}{ c }{$R_2/a^b$}\\
  $[$M31V\,$]$ &           & [mag]         & \multicolumn{2}{ c }{[days]} & \multicolumn{2}{ c }{[deg]} & \multicolumn{2}{ c }{ }   & \multicolumn{2}{ c }{[deg]}    & \multicolumn{2}{ c }{ }         & \multicolumn{2}{ c }{ }        \\
  \hline
J00444528+4128000 & SD & 18.854 & 11.543654 & 0.000211 & 78.0 & 1.3 & \multicolumn{2}{c}{0.0} & \multicolumn{2}{c}{--}   & 0.556 & 0.009 & 0.277 & 0.002 \\
J00435515+4124567 & D  & 19.164 & 6.816191  & 0.000067 & 89.1 & 4.4 & 0.105  &          0.010 & 327.4 & 7.8\hspace{1ex}  & 0.160 & 0.007 & 0.281 & 0.012 \\
J00442326+4127082 & D  & 19.242 & 5.752689  & 0.000037 & 87.3 & 2.4 & 0.189  &          0.019 & 51.3  & 4.7\hspace{1ex}  & 0.215 & 0.017 & 0.267 & 0.025 \\
J00451253+4137261 & SD & 19.366 & 2.358359  & 0.000010 & 75.4 & 3.0 & \multicolumn{2}{c}{0.0} & \multicolumn{2}{c}{--}   & 0.369 & 0.023 & 0.372 & 0.007 \\
J00443799+4129236 & SD & 19.428 & 3.549696  & 0.000012 & 83.3 & 0.6 & \multicolumn{2}{c}{0.0} & \multicolumn{2}{c}{--}   & 0.330 & 0.019 & 0.390 & 0.006 \\
J00442928+4123013 & SD & 19.475 & 3.168969  & 0.000006 & 86.4 & 2.5 & \multicolumn{2}{c}{0.0} & \multicolumn{2}{c}{--}   & 0.329 & 0.014 & 0.403 & 0.005 \\
J00450486+4137291 & SD & 19.536 & 3.094681  & 0.000013 & 75.7 & 3.0 & \multicolumn{2}{c}{0.0} & \multicolumn{2}{c}{--}   & 0.316 & 0.021 & 0.390 & 0.008 \\
J00451973+4145048 & D  & 19.572 & 8.130670  & 0.000087 & 87.2 & 2.4 & 0.252  &          0.027 & 109.1 & 1.8\hspace{1ex}  & 0.216 & 0.014 & 0.149 & 0.012 \\
J00450051+4131393 & D  & 19.756 & 5.211976  & 0.000044 & 81.7 & 1.2 & 0.040  &          0.015 & 120.0 & 14.1\hspace{1ex} & 0.128 & 0.005 & 0.349 & 0.011 \\
J00444914+4152526 & SD & 19.772 & 2.626992  & 0.000010 & 72.5 & 1.5 & \multicolumn{2}{c}{0.0} & \multicolumn{2}{c}{--}   & 0.340 & 0.015 & 0.384 & 0.005 \\
J00442478+4139022 & D  & 20.034 & 4.762310  & 0.000068 & 79.9 & 1.2 & 0.027  &          0.008 & 188.8 & 33.4\hspace{1ex} & 0.105 & 0.003 & 0.333 & 0.014 \\
J00443610+4129194 & SD & 20.038 & 2.048644  & 0.000006 & 70.6 & 1.7 & \multicolumn{2}{c}{0.0} & \multicolumn{2}{c}{--}   & 0.290 & 0.018 & 0.369 & 0.007 \\
J00425907+4136527 & D  & 20.060 & 5.874724  & 0.000069 & 81.0 & 1.1 & 0.163  &          0.016 & 70.8  & 2.2\hspace{1ex}  & 0.344 & 0.013 & 0.157 & 0.011 \\
J00452554+4145035 & SD & 20.090 & 5.009412  & 0.000047 & 82.3 & 3.4 & \multicolumn{2}{c}{0.0} & \multicolumn{2}{c}{--}   & 0.247 & 0.020 & 0.393 & 0.008 \\
J00442722+4136082 & SD & 20.098 & 4.518795  & 0.000016 & 86.1 & 1.7 & \multicolumn{2}{c}{0.0} & \multicolumn{2}{c}{--}   & 0.308 & 0.018 & 0.344 & 0.009 \\
J00453244+4147425 & SD & 20.141 & 2.787856  & 0.000010 & 78.4 & 1.7 & \multicolumn{2}{c}{0.0} & \multicolumn{2}{c}{--}   & 0.238 & 0.012 & 0.342 & 0.007 \\
J00445935+4130454 & SD & 20.166 & 2.668419  & 0.000023 & 68.1 & 1.9 & \multicolumn{2}{c}{0.0} & \multicolumn{2}{c}{--}   & 0.277 & 0.025 & 0.347 & 0.013 \\
J00450537+4133402 & SD & 20.326 & 1.769903  & 0.000005 & 75.4 & 2.5 & \multicolumn{2}{c}{0.0} & \multicolumn{2}{c}{--}   & 0.351 & 0.025 & 0.350 & 0.010 \\
J00432610+4142010 & SD & 20.340 & 4.274429  & 0.000065 & 75.9 & 3.6 & \multicolumn{2}{c}{0.0} & \multicolumn{2}{c}{--}   & 0.206 & 0.027 & 0.413 & 0.014 \\
J00445682+4131109 & D  & 20.360 & 4.207679  & 0.000036 & 87.2 & 4.5 & \multicolumn{2}{c}{0.0} & \multicolumn{2}{c}{--}   & 0.345 & 0.015 & 0.158 & 0.015 \\
J00435082+4121517 & D  & 20.367 & 2.176672  & 0.000011 & 79.6 & 4.1 & \multicolumn{2}{c}{0.0} & \multicolumn{2}{c}{--}   & 0.305 & 0.026 & 0.358 & 0.031 \\
J00425743+4137114 & SD & 20.388 & 1.916302  & 0.000008 & 73.9 & 2.9 & \multicolumn{2}{c}{0.0} & \multicolumn{2}{c}{--}   & 0.364 & 0.019 & 0.379 & 0.006 \\
J00425929+4138169 & D  & 20.450 & 5.591515  & 0.000045 & 85.5 & 4.0 & \multicolumn{2}{c}{0.0} & \multicolumn{2}{c}{--}   & 0.340 & 0.011 & 0.264 & 0.016 \\
J00445011+4128069 & D  & 20.456 & 2.861046  & 0.000009 & 80.2 & 2.4 & \multicolumn{2}{c}{0.0} & \multicolumn{2}{c}{--}   & 0.347 & 0.013 & 0.286 & 0.013 \\
  \hline
 \end{tabular}
 \begin{flushleft}
 $^a$Conf\mbox{}iguration. (D): Detached, (SD): Semi--detached.\\
 $^b R_1/a$ and $R_2/a$ are the relative radii of the components, where $a$ is
the semi-major axis of the system.
 \end{flushleft}

 \begin{tabular}{l r@{$\pm$}l r@{$\pm$}l r@{$\pm$}l r@{$\pm$}l r@{$\pm$}l r@{$\pm$}l r@{$\pm$}l}
  \hline
  \hline
  Identif\mbox{}ier   & \multicolumn{2}{ c }{$T_2^{\rm eff}/T_1^{\rm eff}$} & \multicolumn{2}{ c }{$(L_2/L_1)^B$} & \multicolumn{2}{ c }{$(L_2/L_1)^V$} & \multicolumn{2}{ c }{$(L_2/L_1)^D$} & \multicolumn{2}{ c }{$l_3^B$} & \multicolumn{2}{ c }{$l_3^V$} & \multicolumn{2}{ c }{$l_3^D$} \\
  $[$M31V\,$]$ & \multicolumn{2}{ c }{ }                               & \multicolumn{2}{ c }{ }                 & \multicolumn{2}{ c }{ }                 & \multicolumn{2}{ c }{ }                 &\multicolumn{2}{ c }{ }        & \multicolumn{2}{ c }{ }       & \multicolumn{2}{ c }{ }       \\
  \hline
J00444528+4128000 & 0.453 & 0.033 & 0.0513 & 0.0003 & 0.0575 & 0.0004 & 0.0575  & 0.0004       & \multicolumn{2}{c}{0.0}      & \multicolumn{2}{c}{0.0}     & \multicolumn{2}{c}{0.0}     \\
J00435515+4124567 & 0.911 & 0.018 & 2.69   & 0.09   & 2.70 & 0.08     & \multicolumn{2}{c}{--} & 0.055  & 0.032\hspace{1.3ex} & $-$0.005 & 0.030            & \multicolumn{2}{c}{--}      \\
J00442326+4127082 & 0.890 & 0.027 & 1.25   & 0.13   & 1.27 & 0.14     & 1.27  & 0.15           & 0.329  & 0.039\hspace{1.3ex} & 0.336 & 0.043\hspace{1.3ex} & 0.328 & 0.051\hspace{1.3ex} \\
J00451253+4137261 & 0.885 & 0.020 & 0.80   & 0.08   & 0.82 & 0.09     & 0.82  & 0.12           & 0.285  & 0.056\hspace{1.3ex} & 0.308 & 0.059\hspace{1.3ex} & 0.325 & 0.070               \\
J00443799+4129236 & 0.772 & 0.012 & 0.88   & 0.04   & 0.90 & 0.04     & 0.90  & 0.04           & \multicolumn{2}{c}{0.0}      & \multicolumn{2}{c}{0.0}     & \multicolumn{2}{c}{0.0}     \\
J00442928+4123013 & 0.857 & 0.011 & 1.17   & 0.05   & 1.19 & 0.05     & 1.19  & 0.06           & 0.042  & 0.035\hspace{1.3ex} & 0.027 & 0.032\hspace{1.3ex} & 0.022 & 0.034\hspace{1.3ex} \\
J00450486+4137291 & 0.787 & 0.015 & 1.01   & 0.12   & 1.03 & 0.12     & 1.03  & 0.13           & 0.016  & 0.098\hspace{1.3ex} & $-$0.041 & 0.093            & $-$0.033 & 0.096            \\
J00451973+4145048 & 0.894 & 0.016 & 0.34   & 0.06   & 0.35 & 0.06     & 0.34  & 0.07           & 0.041  & 0.077\hspace{1.3ex} & $-$0.136 & 0.096            & 0.115 & 0.096\hspace{1.3ex} \\
J00450051+4131393 & 0.993 & 0.005 & 7.67   & 0.35   & 7.74 & 0.38     & 7.73  & 1.47           & 0.063  & 0.025\hspace{1.3ex} & 0.069 & 0.024\hspace{1.3ex} & 0.055 & 0.086\hspace{1.3ex} \\
J00444914+4152526 & 0.957 & 0.026 & 1.20   & 0.09   & 1.21 & 0.09     & \multicolumn{2}{c}{--} & 0.003  & 0.046\hspace{1.3ex} & 0.009 & 0.047\hspace{1.3ex} & \multicolumn{2}{c}{--}      \\
J00442478+4139022 & 0.848 & 0.004 & 7.48   & 0.34   & 7.65 & 0.35     & \multicolumn{2}{c}{--} & \multicolumn{2}{c}{0.0}      & \multicolumn{2}{c}{0.0}     & \multicolumn{2}{c}{--}      \\
J00443610+4129194 & 0.946 & 0.020 & 1.51   & 0.18   & 1.52 & 0.18     & 1.52  & 0.22           & 0.067  & 0.034\hspace{1.3ex} & 0.051 & 0.032\hspace{1.3ex} & 0.016 & 0.047\hspace{1.3ex} \\
J00425907+4136527 & 1.046 & 0.021 & 0.22   & 0.02   & 0.22 & 0.02     & 0.22  & 0.02           & 0.053  & 0.017\hspace{1.3ex} & 0.051 & 0.029\hspace{1.3ex} & 0.034 & 0.023\hspace{1.3ex} \\
J00452554+4145035 & 0.810 & 0.018 & 1.72   & 0.13   & 1.78 & 0.13     & 1.78  & 0.27           & 0.169  & 0.014\hspace{1.3ex} & 0.162 & 0.023\hspace{1.3ex} & 0.227 & 0.029\hspace{1.3ex} \\
J00442722+4136082 & 0.747 & 0.008 & 0.72   & 0.05   & 0.76 & 0.05     & 0.76  & 0.06           & 0.109  & 0.009\hspace{1.3ex} & 0.130 & 0.010\hspace{1.3ex} & 0.119 & 0.014\hspace{1.3ex} \\
J00453244+4147425 & 0.728 & 0.009 & 1.14   & 0.11   & 1.20 & 0.12     & 1.20  & 0.17           & 0.036  & 0.022\hspace{1.3ex} & 0.018 & 0.026\hspace{1.3ex} & 0.093 & 0.090\hspace{1.3ex} \\
J00445935+4130454 & 0.857 & 0.063 & 1.22   & 0.24   & 1.25 & 0.25     & \multicolumn{2}{c}{--} & 0.169  & 0.079\hspace{1.3ex} & 0.219 & 0.082\hspace{1.3ex} & \multicolumn{2}{c}{--}      \\
J00450537+4133402 & 0.940 & 0.015 & 0.88   & 0.10   & 0.89 & 0.10     & 0.89  & 0.13           & $-$0.055 & 0.110             & $-$0.027 & 0.102            & $-$0.037 & 0.117            \\
J00432610+4142010 & 0.838 & 0.028 & 3.15   & 0.77   & 3.23 & 0.79     & \multicolumn{2}{c}{--} & 0.243  & 0.043\hspace{1.3ex} & 0.309 & 0.061\hspace{1.3ex} & \multicolumn{2}{c}{--}      \\
J00445682+4131109 & 0.891 & 0.024 & 0.18   & 0.03   & 0.18 & 0.03     & 0.18  & 0.04           & 0.118  & 0.076\hspace{1.3ex} & 0.129 & 0.087\hspace{1.3ex} & 0.153 & 0.117\hspace{1.3ex} \\
J00435082+4121517 & 0.789 & 0.019 & 0.89   & 0.18   & 0.93 & 0.19     & \multicolumn{2}{c}{--} & 0.245  & 0.074\hspace{1.3ex} & 0.211 & 0.091\hspace{1.3ex} & \multicolumn{2}{c}{--}      \\
J00425743+4137114 & 0.921 & 0.014 & 0.94   & 0.11   & 0.96 & 0.11     & \multicolumn{2}{c}{--} & 0.165  & 0.063\hspace{1.3ex} & 0.219 & 0.070\hspace{1.3ex} & \multicolumn{2}{c}{--}      \\
J00425929+4138169 & 0.548 & 0.011 & 0.17   & 0.02   & 0.19 & 0.02     & 0.19  & 0.02           & 0.195  & 0.042\hspace{1.3ex} & 0.203 & 0.049\hspace{1.3ex} & 0.139 & 0.045\hspace{1.3ex} \\
J00445011+4128069 & 0.603 & 0.016 & 0.25   & 0.03   & 0.26 & 0.03     & 0.26  & 0.04           & $-$0.012 & 0.090             & $-$0.004 & 0.080            & 0.001 & 0.105\hspace{1.3ex} \\
  \hline
 \end{tabular}
 }
\end{table*}

\end{document}